\documentclass[10pt,aps,prl,twocolumn,superscriptaddress,showpacs]{revtex4-1}
\usepackage{graphicx}
\usepackage{amssymb}
\usepackage{amsmath}
\usepackage{appendix}

\newcommand{\rf}[1]{(\ref{#1})}
\newcommand{\al}[1]{\begin{aligned}#1\end{aligned}}
\newcommand{\ar}[2]{\begin{array}{#1}#2\end{array}}
\newcommand{\eq}[1]{\begin{equation}#1\end{equation}}

\newcommand{\av}[1]{\langle{#1}\rangle}

\begin{document}

\title{Spin and Orbital Magnetic Response on the Surface of a Topological Insulator}

\author{Yaroslav Tserkovnyak}
\affiliation{Department of Physics and Astronomy, University of California, Los Angeles, California 90095, USA}
\author{D.~A. Pesin}
\affiliation{Department of Physics and Astronomy, University of Utah, Salt Lake City, Utah 84112, USA}
\author{Daniel Loss}
\affiliation{Department of Physics, University of Basel, Klingelbergstrasse 82, CH-4056 Basel, Switzerland}

\begin{abstract}
Coupling of the spin and orbital degrees of freedom on the surface of a strong three-dimensional insulator, on the one hand, and textured magnetic configuration in an adjacent ferromagnetic film, on the other, is studied using a combination of transport and thermodynamic considerations. Expressing exchange coupling between the localized magnetic moments and Dirac electrons in terms of the electrons' out-of-plane orbital and spin magnetizations, we relate the thermodynamic properties of a general ferromagnetic spin texture to the physics in the zeroth Landau level. Persistent currents carried by Dirac electrons endow the magnetic texture with a Dzyaloshinski-Moriya interaction, which exhibits a universal scaling form as a function of electron temperature, chemical potential, and the time-reversal symmetry breaking gap. In addition, the orbital motion of electrons establishes a direct magnetoelectric coupling between the unscreened electric field and local magnetic order, which furnishes complex long-ranged interactions within the magnetic film.
\end{abstract}



\maketitle

\textit{Introduction.|}The magnetoelectric coupling associated with surface states of strong three-dimensional topological insulators (TI's) \cite{pankratovCHA91,*hasanRMP10,*qiRMP11} is at the heart of their phenomenological manifestations. In particular, under appropriate conditions, the electromagnetic response can acquire the form of axion electrodynamics, as embodied by the addition of $\Delta\mathcal{L}=\theta\mathbf{E}\cdot\mathbf{B}$, where $|\theta|=e^2/4\pi\hbar$, to the ordinary Maxwell Lagrangian \cite{wilczekPRL87,*qiPRB08to,*qiSCI09}. $\theta$ is physically manifested by the half-quantized Hall conductance on the TI surface. Finite temperature, bulk conductance, and doping of surface states, however, can easily mask this topological magnetoelectric effect in real systems \cite{pesinPRL13}.

A closely related issue is the magnetoelectric coupling of the TI surface to a proximal ferromagnetic layer \cite{garatePRL10,*nomuraPRB10,tserkovPRL12}, a system that has recently attracted a burst of experimental interest \cite{chenSCI10,*wrayNATP10,*checkelskyNATP12,*fanNATM14,*checkelskyNATP14,*mellnikNAT14}. Strong spin-orbit interaction associated with electron motion on the TI surface couples magnetic order and its spatial texture to the persistent electric currents that are established self-consistently. As a result, the ferromagnetic configuration acquires a Dzyaloshinski-Moriya \cite{zhuPRL11,tserkovPRL12} interaction mediated by the TI electrons, in addition to an out-of-plane anisotropy, which depend on temperature and electrical doping. In this Letter, we formulate an effective theory for the quasiequilibrium spin and orbital response of electrons to the static electromagnetic and spin-exchange fields, in the presence of adjacent ferromagnetic insulator, along with the spin-exchange feedback on the ferromagnetic order. The resulting ground state of the combined ferromagnetic/TI system, in particular, can exhibit a skyrmionic lattice that is tunable by electrical gating or doping.

\textit{Effective theory.|}A minimal effective theory for long-wavelength and low-frequency electromagnetic response of electrons on the ($xy$) surface of a strong three-dimensional TI is provided by the Dirac Hamiltonian \cite{pankratovCHA91}:
\eq{
H_0=v\left(\mathbf{p}+e\mathbf{A}/c\right)\cdot\mathbf{z}\times\hat{\boldsymbol{\sigma}}-e\varphi+\mathfrak{m}\mathbf{B}\cdot\hat{\boldsymbol{\sigma}}\,.
\label{HD}}
Here, $(\varphi,\mathbf{A})$ is the electromagnetic four-vector potential (electric and magnetic fields being given by $\mathbf{E}=-\boldsymbol{\nabla}\varphi-\partial_t\mathbf{A}/c$ and $\mathbf{B}=\boldsymbol{\nabla}\times\mathbf{A}$); $\hat{\boldsymbol{\sigma}}$ is a vector of Pauli matrices, $v$ material-dependent electron velocity, $\mathfrak{m}=(g/2)\mu_B$, $g$ effective $g$-factor, $\mu_B\equiv e\hbar/2mc$ Bohr magneton, $m$ free-electron mass, and $-e<0$ electron charge.

A local axially-symmetric single-particle coupling to an adjacent ferromagnetic film is given by
\eq{
H'=J(n_x\hat{\sigma}_x+n_y\hat{\sigma}_y)+J_\perp n_z\hat{\sigma}_z\,,
\label{Hp}}
where  $\mathbf{n}$ is the directional magnetic order parameter (normalized to $|\mathbf{n}|=1$). Local spin density per unit area in the magnetic film is given by $\mathbf{s}=s\mathbf{n}$, where $s$ is the locally-saturated spin density (assuming that $T\ll T_c$, the Curie temperature). We will suppose the film is insulating, such that its dynamics are fully captured by the precessional motion of $\mathbf{n}(\mathbf{r},t)$ [where $\mathbf{r}=(x,y)$ is the position within the TI surface of interest]. The total Hamiltonian $H=H_0+H'$ describes electronic response to a general electromagnetic field and proximal ferromagnetic layer, as well as encapsulates the reciprocal feedback of the electron spin and charge dynamics.

According to Eq.~\eqref{Hp}, electrons couple to the magnetic film via their spin density $\boldsymbol{\rho}(\mathbf{r})$. The magnetic field $\mathbf{B}$, on the other hand, couples to the orbital current density $\mathbf{j}(\mathbf{r})\perp\mathbf{z}$ through the Peierls substitution and spin density $\boldsymbol{\rho}(\mathbf{r})$ through the Zeeman term. For Dirac electrons, $\mathbf{j}$ and $\boldsymbol{\rho}_\perp\equiv\boldsymbol{\rho}-\mathbf{z}\rho_z$ are helically locked according to $\mathbf{j}=ve\boldsymbol{\rho}\times\mathbf{z}$ [cf. Eq.~\eqref{rhoj} below]. Since, furthermore, the persistent current $\mathbf{j}$ in equilibrium can be recast in terms of the orbital contribution to the out-of-plane magnetization, this magnetization, along with its spin counterpart $\propto\rho_z$, should fully describe quasistationary magnetic response of the TI electrons, both in regard to the magnetic field $\mathbf{B}$ and ferromagnetic magnetization $\mathbf{n}$.

\textit{Spin and orbital magnetizations.|}Let us start by considering a uniform Dirac-electron gas subjected to a constant out-of-plane magnetic field $\mathbf{B}=B\mathbf{z}$ and a uniform time-independent magnetic exchange \rf{Hp}:
\eq{
H=v\left(\mathbf{p}+e\mathbf{A}/c\right)\cdot\mathbf{z}\times\hat{\boldsymbol{\sigma}}+\Delta\hat{\sigma}_z\,,
\label{H}}
where $\Delta\equiv\mathfrak{m}B+J_\perp n_z$ (which controls the gap) and the $J$ terms have been absorbed in the gauge potential $\mathbf{A}$ [which is inconsequential here, but would play a role if we restored the spatiotemporal structure of $\mathbf{n}(\mathbf{r},t)$].

\begin{figure}[pt]
\includegraphics[width=0.75\linewidth,clip=]{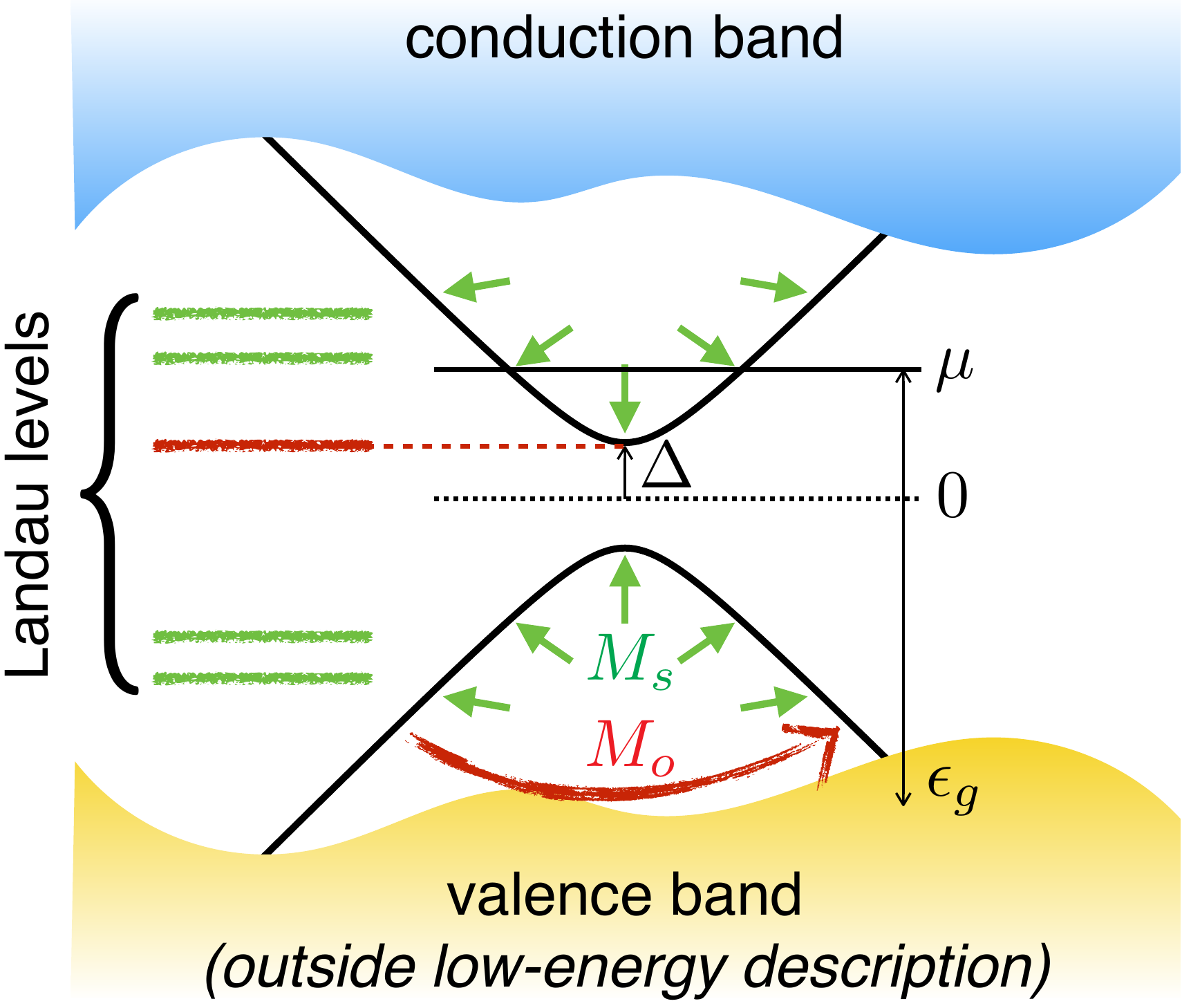}
\caption{Schematic of the low-energy description, which captures contributions to the spin, $M_s$, and orbital, $M_o$, components of the magnetization that stem from the Dirac electrons.}
\label{figS}
\end{figure}

\begin{figure}[pt]
\includegraphics[width=\linewidth,clip=]{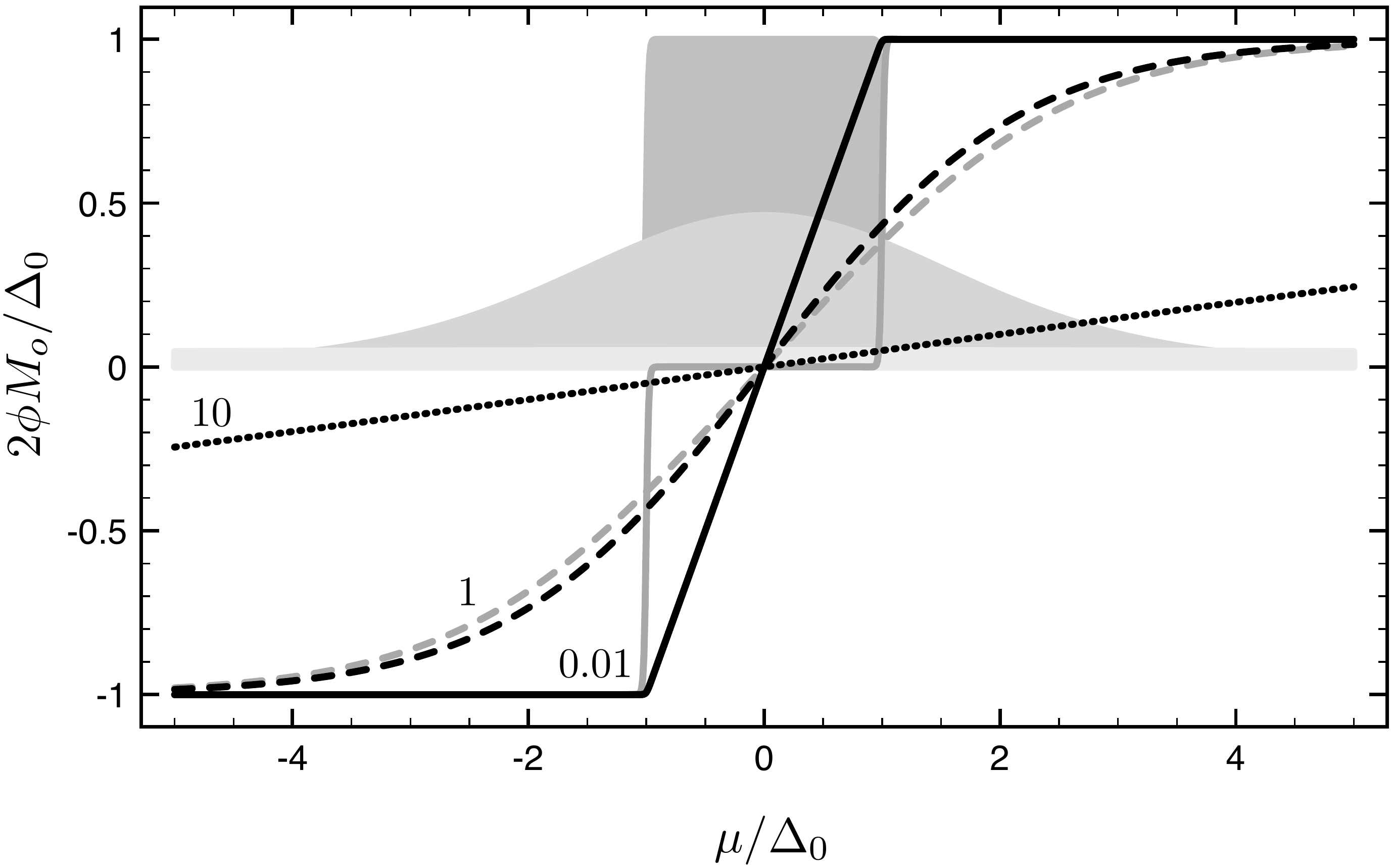}
\caption{Orbital magnetization, Eq.~\eqref{Mo}, is shown by black curves: Solid, dashed, and dotted corresponding to temperatures $k_BT/\Delta_0=0.01$, 1, and 10, respectively, setting $\zeta=0$. The grey curves (with the dotted one essentially overlapping with the black curve) show derivatives $\chi_\Delta$, Eq.~\eqref{MoDo}. The shaded areas are the derivatives $\chi_\mu$, Eq.~\eqref{MoDf}. Note that $\chi_\mu$ vanishes in the extreme limits of both $T\to0$ and $T\to\infty$ for $|\mu/\Delta_0|>1$ \cite{pesinPRL13}.}
\label{figM}
\end{figure}

The total equilibrium out-of-plane magnetization can be calculated thermodynamically, as follows:
\eq{
M=-\partial_B\Omega(\mu,T,B)\,,
}
where $\Omega$ is the (grand-canonical) thermodynamic potential of electrons at chemical potential $\mu$ and ambient temperature $T$, per unit area. To this end, we recall the Landau-level spectrum of the gapped Dirac electrons subjected to a magnetic field \cite{koshinoPRB10}:
\eq{\al{
\epsilon_0=-{\rm sgn}(B)\Delta\,,\,\,\,\epsilon_n={\rm sgn}(n)\sqrt{2(\hbar v/l)^2|n|+\Delta^2}\,,
}}
where $l\equiv\sqrt{\hbar c/e|B|}$ is the magnetic length. $n\in\mathbb{Z}$ is the Landau-level index ($n>0$ corresponding to the electron and $n<0$ valence bands). The degeneracy (per unit area) of each Landau level is given by $\mathcal{N}=1/2\pi l^2$. (See Fig.~\ref{figS} for a schematic of the relevant electronic structure.) This gives for the thermodynamic potential:
\eq{
\Omega(B)=-k_BT\mathcal{N}\sum_n f_n=-k_BT|B|\sum_n f_n(B)/\phi\,,
}
where $f_n\equiv\ln\left[1+e^{\beta(\mu-\epsilon_n)}\right]$, $\beta\equiv(k_BT)^{-1}$, and $\phi\equiv 2\pi\hbar c/e$ is a magnetic flux quantum. When $B=0$,
\eq{\al{
M=&\lim_{B\to0}\frac{\Omega(-B)-\Omega(B)}{2B}
=\frac{k_BT}{2\phi}\ln\frac{1+e^{\beta(\mu+\Delta_0)}}{1+e^{\beta(\mu-\Delta_0)}}\\
&-{\rm sgn}(\Delta_0)\frac{g}{8m\phi}\frac{\Delta_0^2}{v^2}\int_{-y_g}^{y_g}\frac{dy/\varepsilon}{1+e^{\beta(|\Delta_0|\varepsilon-\mu)}}\,,
}\label{M}}
where $\varepsilon\equiv {\rm sgn}(y)\sqrt{1+|y|}$, $\Delta_0\equiv J_\perp n_z$, and $y_g\sim(\epsilon_g/\Delta_0)^2$ is the cutoff for this low-energy theory due to the bulk gap $\epsilon_g\sim\hbar v/a$ ($a$ being the cutoff length scale) of the TI. $M\to M_o+M_s$ consists of two contributions: orbital (Landau-like) magnetization $M_o$, which is governed by the zeroth Landau level, and spin (Pauli-like) magnetization $M_s\propto g$, which is determined by all the other (conduction and valence) Landau levels. The latter corresponds to the spin response induced by the Zeeman term in the Hamiltonian \eqref{HD}, which could also be calculated directly, in the absence of Landau levels. When $\epsilon_g\gg\tilde{\epsilon}$, where $\tilde{\epsilon}=\max(k_BT,|\Delta_0|,|\mu|)$,
\eq{
M_s\sim\Delta_0\frac{g}{4m\phi}\frac{\epsilon_g}{v^2}\,.
\label{Ms}}
The orbital contribution (in the absence of $\mathbf{B}$) is
\eq{\al{
M_o&=\frac{k_BT}{2\phi}\ln\frac{1+e^{\beta(\mu+\Delta_0)}}{1+e^{\beta(\mu-\Delta_0)}}+(\zeta-1)\frac{\Delta_0}{2\phi}\\
&\stackrel{T\to0}{\to}\frac{\Delta_0}{2\phi}\left(\zeta+\left\{\ar{ccc}{1 & , & \mu>|\Delta_0|\\ \mu/|\Delta_0| & , & |\mu|<|\Delta_0|\\ -1 & , & \mu<-|\Delta_0|}\right.\right)\,,
\label{Mo}}}
where we have phenomenologically added a term $\propto(\zeta-1)$ that could stem from energy levels $\epsilon\lesssim-\epsilon_g$, which are beyond our effective theory. (Since, when $\tilde{\epsilon}\ll\epsilon_g$, this contribution should not depend on $\mu$ and can only be weakly dependent on $\Delta_0$, we expanded it to linear order in $\Delta_0$.) The total magnetization $M(B\to0)\propto\Delta_0$, so that, in particular, $M(-\Delta_0)=-M(\Delta_0)$, as should be according to the time-reversal symmetry. Under an additional assumption of the orbital particle-hole symmetry, $\Omega(\mu,B)=\Omega(-\mu,-B)$, we would have $M_o(\mu)=-M_o(-\mu)$, which would imply that $\zeta=0$. In Fig.~\ref{figM}, we plot the orbital magnetization \eqref{Mo}, as a function of chemical potential, at different temperatures.

At first sight, the contribution $\propto\mu$ in Eq.~\rf{Mo} may appear surprising. Indeed, why would a magnetization be modulated by the chemical potential placed inside the gap? This is reconciled by the half-quantized quantum Hall effect and the associated gapless chiral modes at the sample boundary in the $xy$ plane (which, together with the states associated with the opposite TI surface form fully quantized edge states), as follows. When $|\mu/\Delta_0|<1$, $\partial_\mu M_0={\rm sgn}(\Delta_0)/2\phi$, which means that $\partial_\varphi I=g_Q/2$, where $M_o\to I/c$, the charge current at the sample boundary, and $g_Q\equiv e^2/2\pi\hbar$ is the quantum of conductance. (Together with the opposite surface of the 3D TI in the $z$ direction, this would engender a fully quantized integer quantum Hall effect.)

A further insight is afforded by the use of a Maxwell relation:
\eq{
\partial_\mu M=-\partial_\mu(\partial_B\Omega)=\partial_B(-\partial_\mu\Omega)=\partial_Bn\,.
}
Focusing exclusively on the orbital contribution to the magnetization, $M_o$ [and thus the magnetic field entering through the kinetic contribution $\propto v$ in the Hamiltonian \rf{H}], and invoking
\eq{\al{
\left.\partial_B n\right|_{B=0}&=\lim_{B\to0}\left[n(B)-n(-B)\right]/2B\\
&\stackrel{T\to0}{\to}{\rm sgn}(\Delta_0)\mathcal{N}/2|B|={\rm sgn}(\Delta_0)/2\phi\,,
}}
when $|\mu/\Delta_0|<1$ (and zero otherwise), according to the energy flip in the zeroth Landau level. Integrating this over $\mu$, we reproduce Eq.~\rf{Mo}, where $\zeta$ corresponds to the indeterminate constant of integration.

When the chemical potential is in the gap (i.e., $|\mu|<|\Delta_0|$), furthermore, we can invoke the St\v{r}eda formula \cite{stredaJPC82}:
\eq{
g_H=ec\partial_B n\stackrel{B\to0}\to{\rm sgn}(\Delta_0)g_Q/2
}
for the Hall conductance $g_H$ (defined through $\mathbf{j}=g_H\mathbf{z}\times\mathbf{E}$, for the surface current density $\mathbf{j}$) at $T=0$. This completes a logical circle for the relation between the orbital magnetization and the half-quantized Hall response. The pertinent information is thus fully contained in the zeroth Landau level.

\textit{Magnetic-film/TI exchange coupling.|}Equipped with the understanding of how orbital and spin magnetizations form on the TI surface subjected to a proximal magnetic exchange, we return to the general problem of the coupling between Dirac electrons moving along the TI surface, on the one hand, and magnetic moments in the adjacent insulating film, on the other. Integrating over all electrons, the coupling Hamiltonian \rf{Hp} becomes:
\eq{
\mathcal{H}'=\int d^2\mathbf{r}\left[J(n_x\rho_x+n_y\rho_y)+J_\perp n_z\rho_z\right]\,,
\label{Hprho}}
where $\boldsymbol{\rho}(\mathbf{r})$ is the electronic spin density (in units of $\hbar/2$). According to the Dirac Hamiltonian \rf{HD},
\eq{
\boldsymbol{\rho}\times\mathbf{z}=-\frac{c}{ve}\delta_\mathbf{A}H=\frac{\mathbf{j}}{ve}\,,
\label{rhoj}}
where $\mathbf{j}$ is the local (orbital) electric current density. This gives us an operator identity between the planar spin density and electric current. In the following, we will use $\boldsymbol{\rho}$ and $\mathbf{j}$ to denote the expectation values of the corresponding quantities.

In equilibrium (for a static, but now inhomogeneous magnetic film; and at fixed $\mu$ and $T$), we have a one-to-one correspondence between this current and orbital magnetization:
\eq{
\mathbf{j}=c\boldsymbol{\nabla}\times(M_o\mathbf{z})=-c\mathbf{z}\times\boldsymbol{\nabla}M_o\,.
}
In the long-wavelength limit, $M_o$ can be evaluated according to the local $\Delta_0$ and $\mu\to\mu+e\varphi$ [which should replace the chemical potential $\mu$ in Eq.~\rf{Mo}, in the presence of an external scalar potential $\varphi$], which gives:
\eq{
\boldsymbol{\nabla}M_o=\partial_{\Delta_0}M_o\boldsymbol{\nabla}\Delta_0+e\partial_\mu M_o\boldsymbol{\nabla}\varphi\,,
}
where (cf. Fig.~\ref{figM})
\eq{\al{
2\phi\partial_{\Delta_0}M_o&=\zeta+\frac{\sinh(\beta\mu)}{\cosh(\beta\Delta_0)+\cosh(\beta\mu)}\equiv\zeta+\chi_\Delta\\
&\stackrel{T\to0}{\to}\zeta+{\rm sgn}(\mu)\Theta(|\mu/\Delta_0|-1)
\label{MoDo}}}
[here, $\Theta(x)$ is the Heaviside step function, equaling 1 for $x>0$ and zero otherwise],
and
\eq{\al{
2\phi\partial_\mu M_o&=\frac{\sinh(\beta\Delta_0)}{\cosh(\beta\Delta_0)+\cosh(\beta\mu)}\equiv\chi_\mu\\
&\stackrel{T\to0}{\to}{\rm sgn}(\Delta_0)\Theta(1-|\mu/\Delta_0|)\,.
\label{MoDf}}}
Note that $\mu$ entering Eqs.~\rf{MoDo} and \rf{MoDf} is the electrochemical potential relative to the local neutrality point. Combining Eqs.~\rf{rhoj}-\rf{MoDf}, we finally have for the equilibrium planar spin density in the TI:
\eq{\al{
\boldsymbol{\rho}_\parallel&\equiv\rho_x\mathbf{x}+\rho_y\mathbf{y}=\mathbf{z}\times\boldsymbol{\rho}\times\mathbf{z}=\frac{c}{ve}\boldsymbol{\nabla}M_o\\
&=\frac{J_\perp(\zeta+\chi_\Delta)\boldsymbol{\nabla}n_z-e\chi_\mu\mathbf{E}}{4\pi\hbar v}\,.
\label{rhoin0}}}
where $\mathbf{E}=-\boldsymbol{\nabla}\varphi$ is the self-consistent electrostatic field and we assumed that $J_\perp$ is position independent along the interface. We see that at $T\to0$ and $\zeta=0$, the magnetic-texture contribution $\propto\boldsymbol{\nabla}n_z$ to the planar spin density $\boldsymbol{\rho}_\parallel$ vanishes for the electrochemical potential lying inside the gap, while, on the other hand, the electrostatic contribution $\propto\mathbf{E}$ vanishes when the electrochemical potential lies in the continuum of surface states (either conduction, $\mu>|\Delta_0|$, or valence, $\mu<-|\Delta_0|$).

We recall that the contribution to the equilibrium magnetization (and the corresponding current) that is parametrized by $\zeta$ in Eq.~\rf{Mo} is associated with the deeply-lying states that are beyond our low-energy Dirac theory. This means, in particular, that we should not expect the associated states to exchange couple to the magnetic order parameter $\mathbf{n}$ in the same fashion as the Dirac electrons of the effective low-energy description. $\zeta$'s entering in Eqs.~\rf{Mo} and \rf{rhoin0} should thus be viewed as material-dependent phenomenological constants, which may in general be different. $\zeta$ in Eq.~\rf{Mo} is physically relevant for the coupling of the TI surface states to the electromagnetic field, while $\zeta$ in Eq.~\rf{rhoin0} governs the exchange coupling of the TI electrons to the adjacent magnetic film, resulting in the Dzyaloshinski-Moriya energy density $\propto\zeta\mathbf{n}\cdot\boldsymbol{\nabla}n_z$ (which is dictated by symmetry).

The out-of-plane equilibrium spin density $\rho_z$ can be found according to Eq.~\rf{Ms}:
\eq{
\rho_z=-\frac{M_s}{\mathfrak{m}}=-\frac{\xi J_\perp}{2\pi\hbar v a}n_z-\frac{\chi J}{4\pi\hbar v}\boldsymbol{\nabla}\cdot\mathbf{n}\,,
\label{rhoout0}}
where $\xi\sim1$ is a material-dependent parameter and $\chi=\zeta+\chi_\Delta$, which will be shown below Eq.~\rf{Wp}. This last term, which is dictated by structural symmetries, is necessitated by the fact that Eq.~\rf{Ms} was derived for a homogeneous magnetization.

\textit{Magnetic-film dynamics.|}Once the quasistatic spin response of the TI electrons is established, according to Eqs.~\rf{rhoin0} and \rf{rhoout0}, we can write down the equation of motion for the magnetic dynamics, within the Landau-Lifshitz phenomenology \cite{landauBOOKv9,*gilbertIEEEM04}:
\eq{
s(1+\alpha\mathbf{n}\times)\partial_t\mathbf{n}=\mathbf{n}\times\left(\mathbf{H}_{\rm eff}-J\boldsymbol{\rho}_\parallel-J_\perp\rho_z\mathbf{z}\right)\,,
\label{LL}}
where $\mathbf{H}_{\rm eff}\equiv-\delta_\mathbf{n}\Omega_m[\mathbf{n}]$ is the effective field of the isolated magnetic film, due to its intrinsic free-energy functional $\Omega_m[\mathbf{n}]$, $s$ is the film's local spin density, and $\alpha$ is its Gilbert damping constant. If the magnetic dynamics are sufficiently slow, we can use the equilibrium expressions, Eqs.~\rf{rhoin0} and \rf{rhoout0}, for the components of $\boldsymbol{\rho}$: $J\boldsymbol{\rho}_\parallel+J_\perp\rho_z\mathbf{z}=\delta_\mathbf{n}\Omega[\mathbf{n}]$, with $\Omega[\mathbf{n}]=\Omega_0+\Omega'[\mathbf{n}]$ being the electronic contribution to the free energy. The right-hand side of the equation of motion \rf{LL} can thus be written as $\mathbf{n}\times\mathbf{H}^*$, where $\mathbf{H}^*\equiv-\delta_\mathbf{n}(\Omega_m+\Omega')$.

The electronic free energy associated with the exchange coupling \rf{Hprho} can be found (at fixed $\mu$ and $T$) by first integrating $\av{\partial_{J_\perp}\mathcal{H}'}=\int d^2\mathbf{r}\,n_z\rho_z$ over $J_\perp$ from zero up to its physical value, having set $J=0$, and then $\av{\partial_{J}\mathcal{H}'}=\int d^2\mathbf{r}\,\mathbf{n}\cdot\boldsymbol{\rho}_\parallel$ over $J$. This gives:
\eq{\al{
\Omega'[\mathbf{n}]=&\frac{1}{4\pi\hbar v}\int d^2\mathbf{r}\big\{-(\xi J_\perp^2/a)n_z^2\\
&\hspace{5mm}+J\mathbf{n}\cdot\left[J_\perp(\zeta+\chi_\Delta)\boldsymbol{\nabla}n_z-e\chi_\mu\mathbf{E}\right]\big\}\,.
\label{Wp}}}
Noting that $\partial_{J_\perp}\Omega'=\int d^2\mathbf{r}\,n_z\rho_z$, we conclude (after integrating by parts $\int d^2\mathbf{r}\,\mathbf{n}\cdot\boldsymbol{\nabla}n_z\to-\int d^2\mathbf{r}\,n_z\boldsymbol{\nabla}\cdot\mathbf{n}$) that $\chi$ in Eq.~\rf{rhoout0} must indeed be given by $\zeta+\chi_\Delta$. Eq.~\eqref{Wp} is one of the main results of this Letter.

Having effectively integrated electrons out, the TI contribution to the magnetic free-energy density can thus be written as
\eq{
\mathcal{F}'[\mathbf{n}]=-\frac{Kn_z^2}{2}-\frac{\Gamma_{\rm DM}}{2}(n_z\boldsymbol{\nabla}\cdot\mathbf{n}-\mathbf{n}\cdot\boldsymbol{\nabla}n_z)-\Gamma_{\rm ME}\mathbf{E}\cdot\mathbf{n}\,,
\label{F}}
where $K=\xi J_\perp^2/2\pi\hbar va$ is the out-of-plane anisotropy, $\Gamma_{\rm DM}=JJ_\perp(\zeta+\chi_\Delta)/4\pi\hbar v$ is the Dzyaloshinski-Moriya interaction constant, and $\Gamma_{\rm ME}=e\chi_\mu J/4\pi\hbar v$ is a magnetoelectric parameter. $\zeta$ and $\xi$ are nonuniversal dimensionless constants that reflect valence-band physics far away from the Dirac point. $\chi_\Delta$ and $\chi_\mu$, on the other hand, are universal scaling functions (cf. Fig.~\ref{figM}), which describe how the coupling coefficients entering Eq.~\eqref{F} are modulated by the temperature, gap, and electron doping near the Dirac point. Depending on details of the electronic screening, the self-consistent treatment of the term $\propto\mathbf{E}\cdot\mathbf{n}$ may endow the magnetic film with complex long-ranged interactions (that are unrelated to the dipolar coupling).

When the electron-electron interactions are well screened so that $\mathbf{E}\to0$ in Eq.~\eqref{F}, the remaining free-energy density [when complimented with exchange, $A(\partial_i\mathbf{n})^2/2$, and dipolar interactions] has the standard form for a magnetic film with the broken reflection symmetry normal to the film plane \cite{bogdanovJMMM94}. The competition between the Dzyaloshinski-Moriya and exchange interactions, in particular, leads to a spiral ground state with the pitch $\Gamma_{\rm DM}/A$. Subjecting this state to a normal magnetic field $B$ foments two first-order phase transitions, first into a hexagonal crystal of skyrmions, at $B_{c1}$, and then into a ferromagnetic state at $B_{c2}$, where $B_{c1}\lesssim B_{c2}\sim\Gamma_{\rm DM}^2/AM$ ($M$ being the saturation magnetization of the film), at low temperatures \cite{bogdanovJMMM94,hanPRB10}. The intermediate-field skyrmionic phase is particularly interesting from the point of view of anomalous magnon transport \cite{hoogdalemPRB13} and thermally-driven collective magnetic dynamics \cite{mochizukiNATM14}.

One of the intriguing conclusions of this work is that these phase transitions can be controlled electrically through the contribution $\propto\chi_\Delta$ to $\Gamma_{\rm DM}$. In particular, at $T\to0$, $\chi_\Delta\to{\rm sgn}(\mu)\Theta(|\mu/\Delta_0|-1)$ [cf. Eq.~\eqref{MoDo}], which becomes $\pm1$ at sufficiently large (low) electrochemical potential $\mu$, irrespective of the magnetic texture $\mathbf{n}(\mathbf{r})$ that controls $\Delta_0\equiv J_\perp n_z$. By electrically gating the TI surface, we could thus modulate $\Gamma_{\rm DM}$, which, in turn, governs the equilibrium magnetic structure. Since, furthermore, the universal scaling function $\chi_\Delta(\Delta,\mu,T)$ reflects the Dirac nature of electrons on the TI surface, measuring $\Gamma_{\rm DM}\propto\zeta+\chi_\Delta$ as a function of the electron density and temperature could offer direct magnetoelectric signatures of the topological character of the surface.

\begin{acknowledgments}
This work was supported by FAME (an SRC STARnet center sponsored by MARCO and DARPA) and in part by the US DOE-BES under Award No.~DE-SC0012190 and the ARO under Contract No.~911NF-14-1-0016 (Y.T.), Swiss NSF and NCCR QSIT (D.L.).
\end{acknowledgments}

\end{document}